\documentclass{article}
\usepackage[utf8]{inputenc}
\usepackage{authblk}
\usepackage{setspace}
\usepackage{braket}
\usepackage[margin=1.25in]{geometry}
\usepackage{graphicx}
\graphicspath{ {./figures/} }
\usepackage{subcaption}
\usepackage{amsmath}
\usepackage{physics}
\usepackage{booktabs}
\usepackage{lineno}
\usepackage{xcolor}
\usepackage{tabularx}
\usepackage{array}
\usepackage{cite}

\title{Experimental Efficient Source-Independent Quantum Conference Key Agreement}

\author[1,2$\dag$]{Wen-Ji Hua}
\author[1,2$\dag$]{Yi-Ran Xiao}
\author[1,2]{Yu Bao}
\author[2,1,3,4*]{Hua-Lei Yin}
\author[1*]{Zeng-Bing Chen}

\affil[1]{National Laboratory of Solid State Microstructures and School of Physics, Collaborative Innovation Center of Advanced Microstructures, Nanjing University, Nanjing 210093, China.}
\affil[2]{School of Physics and Key Laboratory of Quantum State Construction and Manipulation (Ministry of Education), Renmin University of China, Beijing 100872, China.}
\affil[3]{Beijing Academy of Quantum Information Sciences, Beijing 100193, China.}
\affil[4]{Yunnan Key Laboratory for Quantum Information, Yunnan University, Kunming 650091, China}
\affil[*]{Address correspondence to: hlyin@ruc.edu.cn(H.-L.Y); zbchen@nju.edu.cn (Z.-B.C.)}
\affil[$\dag$]{These authors contributed equally to this work.}

\date{}

\begin{document}

\maketitle

\begin{abstract}
Multipartite entanglement enables secure group key distribution among multiple users while providing immunity against hacking attacks targeting source devices, thereby realizing source-independent quantum conference key agreement (SI-QCKA). However, previous experimental demonstrations of SI-QCKA have encountered substantial technical challenges, primarily due to the low efficiency and scalability limitations inherent in the generation and distribution of multipartite entanglement. Here, we experimentally demonstrate a scalable and efficient SI-QCKA protocol using polarization-entangled photon pairs in a three-user star network, where Greenberger-Horne-Zeilinger correlations are realized via a post-matching method. We achieve a secure group key rate of $2.11 \times 10^{4}$ bits/s under the single-user channel transmission of 1.64 $\times$ $10^{-1}$ in a symmetric channel loss network. Additionally, we conduct six sets of experiments to investigate the impact of varying channel transmission and random basis selection probabilities on secure key rates. Our work establishes an efficient pathway for SI-QCKA and demonstrates potential scalability for future large-scale multi-user quantum networks.
\end{abstract}

\section{Introduction}
Quantum networks are envisioned to facilitate global-scale secure quantum communication between any users through state-of-the-art quantum technologies. To date, several categories of quantum network configurations have been developed~\cite{wehner2018quantum,azuma2023quantum,li2024asynchronous}, supporting various applications including distributed quantum sensing and metrology~\cite{guo2020distributed,zhao2021field,xu2024integrated,zhou2023quantum}, distributed quantum computing~\cite{zhong2021deterministic,zhou2022experimental,zhang2024quantum} and quantum communication~\cite{cao2024experimentala,chen2021integrated,weng2023beating,jing2024experimental,cao2023realization,yang2025300}, which are fundamentally unattainable through classical networks. Quantum cryptography is essential for enabling secure communication in quantum networks~\cite{yin2023experimental,bozzio2024quantum,pan2024evolution}. As a representative application of quantum cryptography, quantum key distribution (QKD) has attracted extensive attention due to its information-theoretic security~\cite{xu2020securea,pirandola2020advances}. Remarkable progress has been achieved in both theoretical frameworks and experimental implementations of QKD~\cite{lo2012measurementdeviceindependent,lucamarini2018overcoming,xie2022breaking,zeng2022mode,ying2025passivestate}, establishing a robust foundation for its practical applications~\cite{wang2024experimental,gu2022experimentala,guo2025discretemodulated,liu2021homodyne,pan2025simultaneous,ding2025quantum,sheng2022onestep,fang2025fast,Shao2025intergration}. To extend practical applications of QKD, point-to-point protocols must evolve into multi-user configurations. Quantum conference key agreement (QCKA) is a notable application~\cite{fu2015longdistance,pickston2023conferencea,grasselli2019conference,murta2020quantum}, enabling the distribution of secure keys among multiple users.  Additionally, more efficient protocols leveraging multipartite entanglement have been developed to reduce resource requirements~\cite{li2023allphotonica,epping2017multipartitea,xie2024multi,wallnofer2019multipartite,miguel-ramiro2023optimizeda,kuzmin2019scalable,lu2025repeaterlike}.

Despite extensive theoretical proposals of QCKA protocols~\cite{epping2017multipartitea,fu2015longdistance,grasselli2019conference,hahn2020anonymousa,cao2021coherent,li2021finitekey,grasselli2018finitekey,ribeiro2018fully,cao2021high,zhao2020phasematching,li2023breaking}, technical difficulties have constrained experimental implementations to primarily focus on multipartite entanglement distribution schemes~\cite{pickston2023conferencea,ho2022entanglementbased,cai2018multipartite,proietti2021experimental}. A representative experimental demonstration based on the N-BB84 protocol~\cite{grasselli2018finitekey}, employing four-photon Greenberger-Horne-Zeilinger (GHZ) entangled states, has effectively demonstrated secure key distribution through a 50 km fiber~\cite{proietti2021experimental}. However, the experimental requirements for high-intensity entanglement sources and long-distance communication channels hinder its practical implementation~\cite{webb2024experimental,pickston2023conferencea,jons2017bright,gaertner2007experimental,tittel2001experimental,erven2014experimental,uppu2021quantumdotbased}. 

To circumvent the obstacles, measurement-device-independent quantum conference key agreement (MDI-QCKA) protocols have been developed~\cite{fu2015longdistance}. MDI-QCKA establishes quantum correlations among users by postselecting GHZ entangled states through a multiphoton interferometer~\cite{pan1998greenbergerhornezeilingerstate}, thereby permitting measurement devices to be controlled by an untrusted party. The MDI-QCKA experiment has been successfully implemented~\cite{yang2024experimental,du2025experimental}. However, the high demands for time synchronization and pulse indistinguishability introduce considerable practical difficulties. Moreover, the complexity of the GHZ analyzer grows rapidly as the number of users increases. Therefore, the realization of QCKA in practical quantum networks remains a major challenge.

Here, we experimentally demonstrate an efficient multi-user source-independent QCKA protocol~\cite{bao2024efficient} based on Bell state distribution. We note that entanglement-based protocols are naturally source-independent, which guarantees the system is secure against all security loopholes related to source imperfections or active hacking attacks targeting the source device~\cite{ma2007quantuma,koashi2003securea}. The protocol establishes multipartite quantum correlations via the post-matching method~\cite{lu2021efficient}, thus avoiding the need for multipartite entangled state generation. By employing polarization-entangled photon pair sources, we have successfully implemented a tripartite QCKA system, where fidelity and visibility of the bipartite entanglement between each pair of users reach up to 97\% and 96\%, respectively. We conduct six sets of tests under varying channel transmission and basis selection probabilities, attaining a high key rate of $2.11 \times 10^4$ bit/s at the channel transmission of $1.64 \times10^{-1}$ and a $Z$-basis selection probability of 0.9. Furthermore, the implementation can be integrated into quantum networks, such as the fully connected QKD network architecture~\cite{huang2024sixteenuser}, by leveraging dense wavelength division multiplexing technology. As user numbers scale, the system requires only additional detection devices, ensuring flexibility for user adjustments. Therefore, this approach offers a resource-efficient, technically feasible, and scalable scheme for future large-scale quantum multiparty networks.

\section{Results}
\subsection{Protocol description}
In Fig.~\ref{fig1}, we present a schematic overview of the source-independent QCKA protocol's workflow, emphasizing the distribution of Bell pairs and the establishment of GHZ correlation. Unlike many previous protocols utilizing GHZ states, the source-independent QCKA protocol~\cite{bao2024efficient} enables multiple users to acquire secure keys using entangled photon pair sources. The required maximally entangled Bell states are expressed as $\ket{\Phi^+}=(\ket{00}+\ket{11})/\sqrt{2}$ in the $Z$ basis and $\ket{\Phi^+}=(\ket{++}+\ket{--})/\sqrt{2}$ in the $X$ basis. We note that the experimentally generated $\ket{\Psi^-}$ state, used for its higher visibility, can be transformed into the required $\ket{\Phi^+}$ state. This transformation, which corresponds to Alice applying an $X_A Z_A$ local operation, is accounted for computationally in classical post-processing, as will be detailed in Section 2.4. Here, $\ket{0}$ and $\ket{1}$ represent the eigenstates of the $Z$ basis, while $\ket{+}$ and $\ket{-}$ denote the eigenstates of the $X$ basis, with $\ket{+}= (\ket{0}+\ket{1})/\sqrt{2}$ and $\ket{-}=(\ket{0}-\ket{1})/\sqrt{2}$. For $n$ users, denoted as Alice, $\text{Bob}_1$, $\text{Bob}_2$,..., $\text{Bob}_{(n-1)}$, the key steps of the protocol are outlined as follows:

\begin{enumerate}
    \item Entanglement distribution. In each round, an untrusted central node (Eve) generates $ n-1 $ polarization-entangled photon pairs in the state $ \ket{\Phi^{+}} = (\ket{00} + \ket{11})/\sqrt{2} $. These pairs are distributed with one photon sent to Alice and the other to each user $\text{Bob}_i$ ($i= 1, 2, \dots, n-1$).

    \item Basis selection and measurement. Each user independently measures the photons in either the $ Z $ basis with probability $ p_z $ or the $ X $ basis with probability $ 1-p_z $. Measurement results are recorded as valid events only when Alice and $\text{Bob}_i$ select the same measurement basis and simultaneously detect photons.

    \item Post-matching and key correlation. Users publicly announce their basis choices. Valid events are grouped according to the chosen basis and sorted by measurement time. The measurement results are ordered and labeled as $ a^{iz}_j $ ($ a^{ix}_j $) for Alice and $ b^{iz}_j $ ($ b^{ix}_j $) for $\text{Bob}_i$, where $ i $ corresponds to $\text{Bob}_i$ and $ j $ denotes the time sequence.
    
    For $ Z $-basis measurement results: Alice computes the parity bits $ c_j^{iz} = a_j^{1z} \oplus a_j^{iz} $ and broadcasts them. Each $\text{Bob}_i$ derives the correlated keys via $ b_j^{iz'} = c_j^{iz} \oplus b_j^{iz} $, thereby obtaining bit strings with GHZ correlation $ a_j^{1z} = b_j^{1z'} = \dots = b_j^{(n-1)z'} $. 
    
    For $ X $-basis measurement results: Alice computes $ a_j^{1x'} = \bigoplus_{i} a_j^{ix} $. The expected outcome is $ a_j^{1x'} = \bigoplus_{i} b_j^{ix} $, which can be verified to establish bounds on potential eavesdropping.
    
\begin{figure}[!htbp]
    \centering
    \includegraphics[width=0.8\columnwidth]{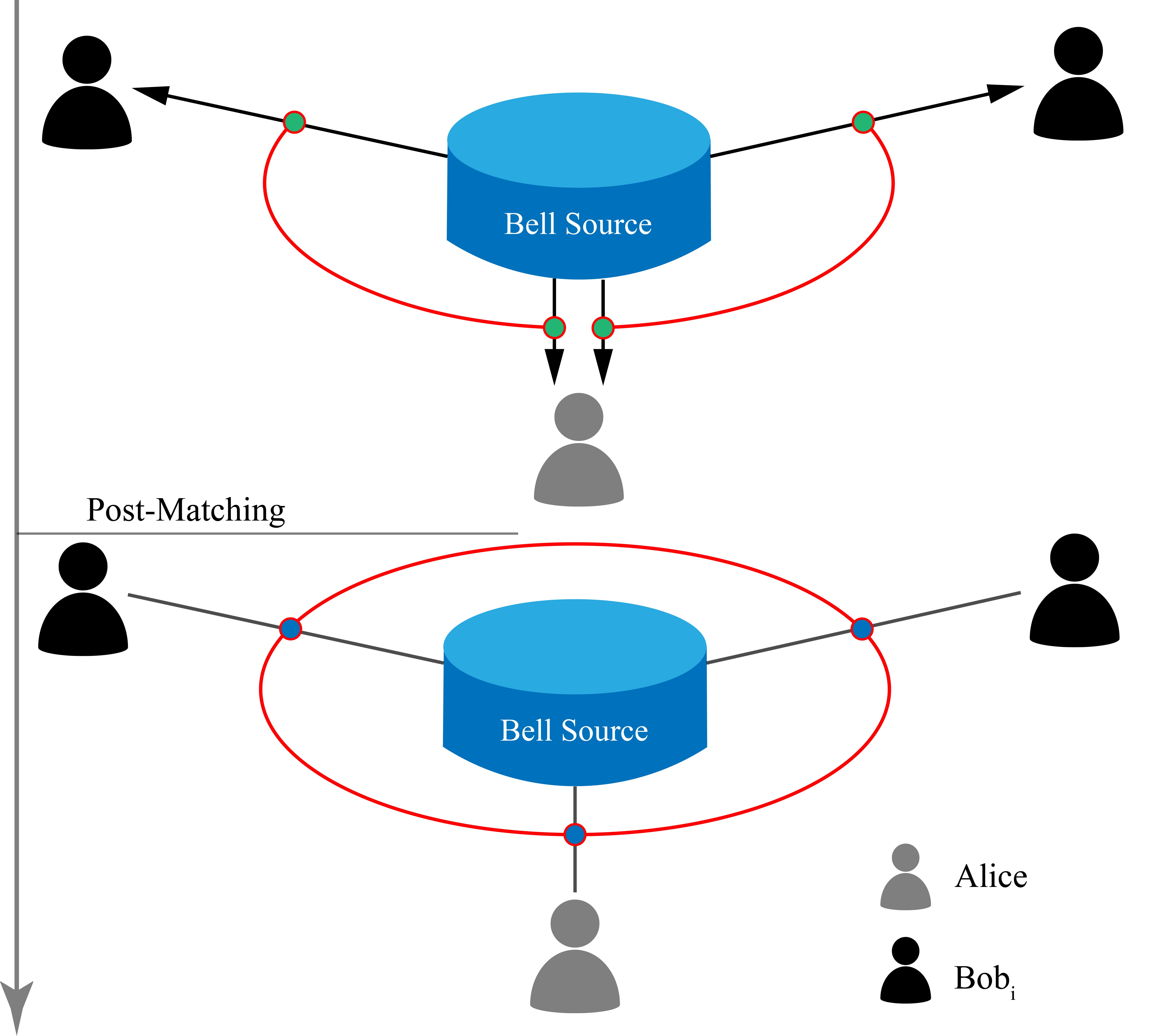}
    \caption{Schematic of the three-party source-independent QCKA protocol (Alice, $\text{Bob}_1$ and $\text{Bob}_2$). The central node functions as a Bell source, generating entangled Bell pairs that are distributed to Alice-$\text{Bob}_1$ and Alice-$\text{Bob}_2$. Subsequently, the users perform random measurements on the received photons, choosing either the $Z$ basis or the $X$ basis. The event in which a pair is measured by Alice and $\text{Bob}_1$ (or $\text{Bob}_2$) in the same basis will be recorded as a valid event. After processing the measurement results of the valid events using the post-matching method, we can establish GHZ correlation, from which raw key strings can be generated.}
    \label{fig1}
\end{figure}

\item Error correction. The error rate in the $X$ basis is evaluated to quantify the potential information leakage to eavesdroppers. Subsequently, error correction is performed on the raw keys obtained from $Z$-basis measurements to ensure identical bit strings between the communicating users.

\item Privacy amplification. To reduce the residual information potentially acquired by an adversary, privacy amplification is applied to the keys. This process compresses the key length according to the estimated information leakage and generates the final secure keys.
\end{enumerate}

\begin{figure*}[!htbp]
    \centering
    \includegraphics[width=\linewidth]{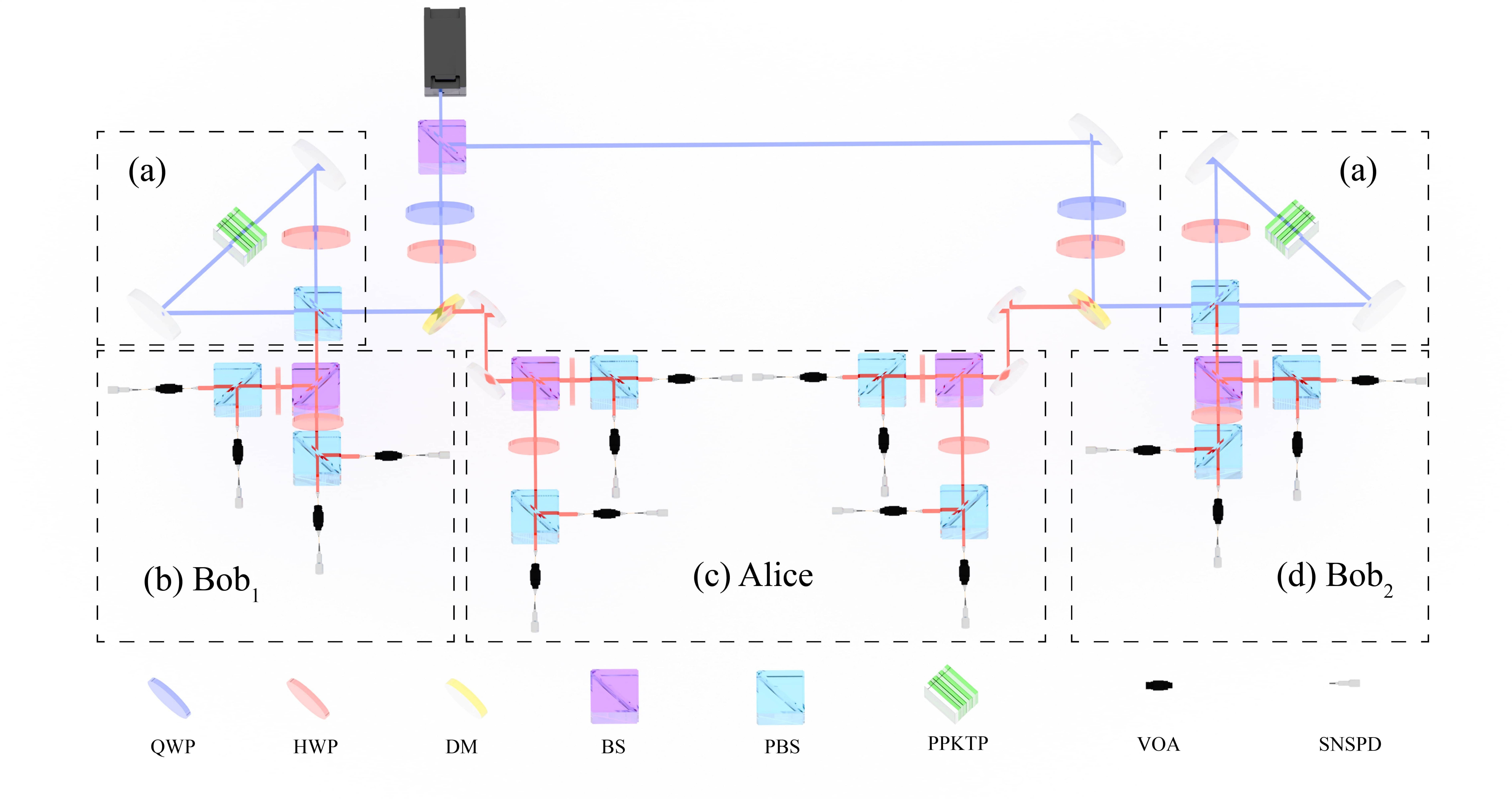}
    \caption{Schematic of the QCKA experimental setup. (a) Sagnac loops. Laser pulses centered at a wavelength of 780 nm are injected into Sagnac loops, where polarization-entangled photon pairs at a wavelength of 1560 nm are generated via type-II spontaneous parametric down-conversion in a periodically poled potassium titanyl phosphate (PPKTP) crystal. The half wave plates (HWPs) in the Sagnac loops are operational at both 780 nm and 1560 nm wavelengths. Entangled photon pairs are sequentially allocated to Alice, $\text{Bob}_1$ and $\text{Bob}_2$ for polarization projection measurements. (b), (c) and (d) The single-photon polarization measurements of the users. In the measurement sites, beam splitters (BSs) are utilized for random basis selection. The HWPs following the BS cooperate with the polarizing beam splitter (PBS) to perform projection measurements. The HWPs are set at $0^\circ$/$45^\circ$ and $22.5^\circ$/$67.5^\circ$  for measurements in the horizontal/vertical (H/V) and diagonal/anti-diagonal (D/A) bases, respectively. Photons are then measured in different bases and detected by distinct SNSPDs. QWP, quarter-wave plate; DM, dichroic mirror. }
    \label{fig2}
\end{figure*}

\subsection{Experimental setup}

The experimental setup is illustrated in Fig.~\ref{fig2}, involving three users: Alice, $\text{Bob}_1$ and $\text{Bob}_2$. A picosecond pulse laser at 780 nm with a repetition frequency of 96.7 MHz is first split by a 50:50 beam splitter (BS) and subsequently injected into two Sagnac loops, generating two polarization-entangled photon pairs simultaneously at a wavelength of 1560 nm. The Sagnac interferometer splits the 780 nm pump pulse into two counter-propagating beams that traverse the same PPKTP crystal. Because these two paths are inherently phase-stable and perfectly overlapped spatially, the photon pairs generated from both directions are fundamentally indistinguishable. This indistinguishability results in a coherent superposition of the two processes, which is the key mechanism for creating the high-fidelity polarization-entangled state.  The state of the entangled photon pairs is prepared as $\ket{\Psi^-} = (\ket{HV}-\ket{VH})/\sqrt{2}$ to achieve higher visibility. Here, $\ket{H}$ and $\ket{V}$ denote horizontally and vertically polarized photons, respectively. 

Subsequently, the photon pairs are split and delivered to Alice, $\text{Bob}_1$ and $\text{Bob}_2$ respectively, where projection measurements are performed. Within the two separate photon pairs, Alice receives one photon from each entangled pair, while the remaining two photons are allocated to two distinct users, thereby establishing unique polarization-entangled correlations between Alice and each individual user. Alice has twice the devices of other users to measure two photons  simultaneously. Notably, all users employ BSs with identical splitting ratios, and the probability of basis selection is determined by the ratio of the BS. Photons traveling along the two arms of the BS are measured in the $Z$ basis and the $X$ basis, respectively. The variable optical attenuators (VOA) are utilized to evaluate the system's performance under varying channel transmission. After passing through the VOAs, the photons are detected by a superconducting nanowire single-photon detector (SNSPD). All the SNSPDs are connected to a time-to-digital converter to record the arrival times of the detected photons. The average dark count rates of the SNSPDs in our experiment are below 20 Hz, with an average detection efficiency of 83\%. An event in which Alice and $\text{Bob}_1$ (or $\text{Bob}_2$) simultaneously detect a photon under the same basis within a time window of 5.16 ns is recorded as a correlation count, corresponding to a valid event in the protocol.

To maximize the recording of valid events, the photons within the same entangled pair must arrive at the SNSPDs with minimal time difference. Consequently, the optical paths traveled by the photons are carefully adjusted to be as equal as possible. Additionally, appropriate time delays are applied to the signals from different SNSPDs to compensate for subtle differences in the optical path lengths. The introduced time delays are kept minimal, with their values optimized to maximize the coincidence count rates. The coincidence counts are then utilized to extract the secure keys and estimate information leakage following the protocol.

\subsection{Performance}

\begin{figure*}[htbp]
    \centering
    \includegraphics[width=\linewidth]{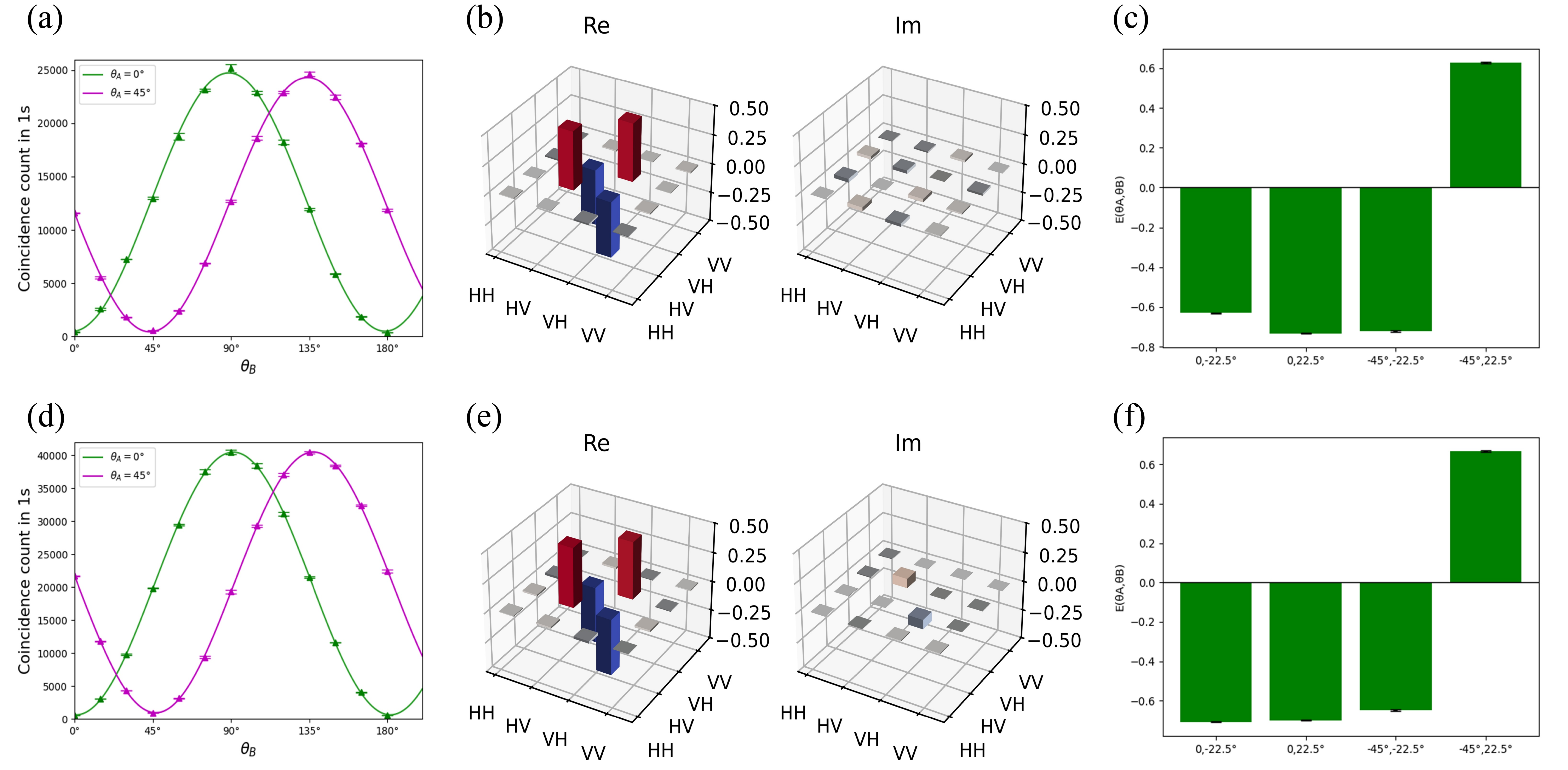}
    \caption{The characterization parameters of the two entangled photon pairs. (a)-(c) correspond to Alice and $\text{Bob}_1$, and (d)-(f) correspond to Alice and $\text{Bob}_2$. (a) and (d) Two-photon interference fringes as functions of the polarizer angles of Alice and $\text{Bob}_1$ (or $\text{Bob}_2$) under two different phase bases. Alice sets $\theta_A$ to $0^{\circ}$ (green points, fitted with green line) and $45^{\circ}$ (magenta points, fitted with magenta line), while $\text{Bob}_i$ sweeps $\theta_{B_i}$ from $0^{\circ}$ to $180^{\circ}$. (b) and (e) Quantum state tomography results. The real and imaginary parts of the density matrix are depicted with respect to $\ket{\Psi^-} = (\ket{HV}-\ket{VH})/\sqrt{2}$. (c) and (f) The four expectation values used to calculate the S parameter of the CHSH inequality. (c) $E_{AB_1}(0^{\circ},-22.5^{\circ}) = -0.631 \pm 0.004$, $E_{AB_1}(0^{\circ},22.5^{\circ}) = -0.733 \pm 0.003$, $E_{AB_1}(-45^{\circ},-22.5^{\circ}) = -0.723 \pm 0.003$, $E_{AB_1}(-45^{\circ},22.5^{\circ}) = 0.627 \pm 0.004$. (f) $E_{AB_2}(0^{\circ},-22.5^{\circ}) = -0.708 \pm 0.003$, $E_{AB_2}(0^{\circ},22.5^{\circ}) = -0.700 \pm 0.003$, $E_{AB_2}(-45^{\circ},-22.5^{\circ}) = -0.647 \pm 0.003$, $E_{AB_2}(-45^{\circ},22.5^{\circ}) = 0.666 \pm 0.003$.}
    \label{fig3}
\end{figure*}

It's vital to ensure that high-quality entanglement is generated and distributed to all users. Therefore, we perform a series of experiments to evaluate the performance of the experimental system. During these experiments, all the VOAs are set to 0 dB, while the total channel transmission excluding the detector efficiency in the system between the entanglement source (Eve) and each user (Alice, $\text{Bob}_1$ and $\text{Bob}_2$) is maintained at approximately 1.64 $\times$ $10^{-1}$. Additionally, the BSs employed for random basis selection are not plugged in the setup during the experiments and the HWPs in the measurement sites are set at the required angles based on the experimental demands. To minimize the influence of the extinction ratio difference between the transmission and reflection paths of the PBS before VOAs, only the transmitted photons are detected and recorded. The average down-converted mean photon number $\mu$ of the two entangled photon pairs is about 0.023 per pulse, remaining consistent throughout the demonstration experiment. In the characterization experiments, the coincidence count rates of both pairs of users exceed $10^4$ per second, with the coincidence count to accidental coincidence count ratios surpassing 50, which indicates that each pair of users successfully shares a pair of entangled photons with high fidelity.

Furthermore, the quality of the polarization entanglement shared by the users can be quantitatively assessed through the visibility and fidelity of the entangled photon pairs. To measure the visibility of the photon pairs, we perform a two-photon interference experiment. Fig.~\ref{fig3}(a) illustrates the photon interference fringes between Alice and $\text{Bob}_1$, while Fig.~\ref{fig3}(d) depicts those between Alice and $\text{Bob}_2$. $\theta_A$ denotes the polarization angle of Alice, and $\theta_{B_i}$ denotes the polarization angle of $\text{Bob}_i$. We fix $\theta_A$ at $0^{\circ}$ and $45^{\circ}$ while sweeping $\theta_{B_i}$ from $0^{\circ}$ to $180^{\circ}$ to measure the coincidence counts $N$. The average visibility is calculated by $V = (N_{\mathrm{max}} - N_{\mathrm{min}})/(N_{\mathrm{max}} + N_{\mathrm{min}})$. For Alice and $\text{Bob}_1$, the average visibility is $ V_{AB_1} = 96.43\% \pm 0.26\% $, and for Alice and $\text{Bob}_2$, it is $ V_{AB_2} = 96.12\% \pm 0.76\% $. Both of them surpass the $1/\sqrt{2}$ visibility threshold required for the violation of the Clauser–Horne–Shimony–Holt (CHSH) inequality~\cite{clauser1969proposeda}. 

Then, we plug in an additional QWP before each HWP at all the measurement sites of the users to perform quantum state tomography and calculate the fidelity of the entangled photons with the maximum likelihood estimation of the density matrix~\cite{james2001measurement}. The reconstruction of the density matrix is detailed in Section 4.3. The density matrix of the photon pairs, estimated based on the experimental data, is shown in Fig.~\ref{fig3}. For Alice and $\text{Bob}_1$, the average fidelity with respect to the Bell state $\ket{\Psi^-} = (\ket{HV}-\ket{VH})/\sqrt{2}$ is $F_{AB_1} = 97.08\% \pm 0.03\%$, and for Alice and $\text{Bob}_2$, it is $F_{AB_2} = 96.98\% \pm 0.06\%$, revealing the high quality of the entanglement shared by the users. 

Finally, we calculate the four expectation values required for computing the CHSH inequality parameter $ S $ by measuring the coincidence counts under four distinct combinations of wave plate angles~\cite{clauser1969proposeda}, i.e., $S = E(0^{\circ},-22.5^{\circ}) + E(0^{\circ},22.5^{\circ}) + E(-45^{\circ},-22.5^{\circ}) - E(-45^{\circ},22.5^{\circ})$. For Alice and $\text{Bob}_1$, $\abs{S_{AB_1}} = 2.714 \pm 0.006$, and for Alice and $\text{Bob}_2$, $\abs{S_{AB_2}} = 2.720 \pm 0.005$, both of which are close to the theoretical upper bound of $2\sqrt{2}$. The detailed values are illustrated in Fig.~\ref{fig3}.

\begin{table*}[!ht]
    \caption{Summary of experimental data. $10^{11}$ pulses are sent in total. $\eta$ denotes the channel transmission excluding the detector efficiency between central node Eve and each user. The measured channel transmission values for all channels are approximately identical, i.e., $\eta=\eta_{EA}=\eta_{EB_1}=\eta_{EB_2}$, where $\eta_{EA}$ is the channel transmission from Eve to Alice, and $\eta_{EB_i}$ is the channel transmission from Eve to $\text{Bob}_i$. $E_{AB1}^X$, $E_{AB1}^Z$ denote the experimental quantum bit error rates between Alice and $\text{Bob}_1$ in $X$ basis and $Z$ basis, respectively, and $E_{AB2}^X$, $E_{AB2}^Z$ denote the corresponding quantum bit error rates between Alice and $\text{Bob}_2$. $R_{\mathrm{QCKA}}$ is the experimental conference secure key rate.}
    \begin{tabular*}{\hsize}{@{}@{\extracolsep{\fill}}lllllll@{} }
    \toprule
    $p_z$ &  $\eta$ & $E_{AB1}^X$(\%) & $E_{AB2}^X$(\%) & $E_{AB1}^Z$(\%) & $E_{AB2}^Z$(\%) & $R_{\mathrm{QCKA}}$(bit/s) \\
    \midrule
       0.9  & 1.64 $\times$ $10^{-1}$ & 1.46 & 1.09 & 2.54 & 2.66 & 2.11 $\times$ $10^{4}$ \\
           & 8.24 $\times$ $10^{-2}$ & 1.50 & 1.19 & 3.02 & 2.35 & 4.72 $\times$ $10^{3}$\\
           & 5.20 $\times$ $10^{-2}$ & 1.57 & 1.24 & 3.24 & 2.34 & 1.72 $\times$ $10^{3}$ \\
           \midrule
       0.5 & 1.64 $\times$ $10^{-1}$& 1.99 & 1.91  & 1.79  & 1.92 & 6.25 $\times$ $10^{3}$ \\
           & 8.24 $\times$ $10^{-2}$ & 2.04 & 2.18 & 1.99 & 2.08 & 1.55 $\times$ $10^{3}$\\
           & 5.20 $\times$ $10^{-2}$ & 2.11 & 2.10 & 2.00 & 2.08 & 6.31 $\times$ $10^{2}$\\
    \bottomrule
    \end{tabular*}
    \label{tab:1}
\end{table*}

\subsection{Experimental results}
We experimentally demonstrate the tripartite source-independent QCKA protocol utilizing the established polarization-entangled photon pair source. Through applying the post-matching method to the measurement results, users can generate the raw keys and estimate the information leakage based on the coincidence counts. The length of the secure key is given by~\cite{li2023breaking,grasselli2018finitekey,yin2020tight}:
\begin{equation}
\begin{aligned}
     L_{\mathrm{QCKA}} & = n_Z\left\{1-H(\phi^Z)- f \max\limits_i \left[H(E_{AB_i}^Z)\right]\right\} -\log_2 \frac{2(n-1)}{\epsilon_{\mathrm{cor}}} - 2 \log_2 \frac{1}{2\epsilon_{\mathrm{sec}}}
\end{aligned}
\label{eq:1}
\end{equation}
where $n_Z$ is the number of entangled photon pairs detected in the $Z$ basis, $H(x) = -x\log_2(x) - (1-x) \log_2 (1-x) $ represents the binary Shannon entropy and  $\phi^Z$ is the upper limit of the phase error rate in the $Z$ basis accounting for statistical fluctuations. $f$ is the error correction efficiency, $E_{AB_i}^Z$ is the marginal error rates between Alice and the corresponding $\text{Bob}_i$, $n$ is the number of users, $\epsilon_{\mathrm{cor}}$ is the failure probability of error verification and $\epsilon_{\mathrm{sec}}$ is the failure probability of privacy amplification. In the experiment, $n=3$, and $n_Z$, $\phi^Z$ and $E_{AB_i}^Z$ are calculated from the experiment data~\cite{bao2024efficient}. The remaining parameters used are provided in Sec.~\ref{sec:4.1}.

Given that the quantum bit error rates (QBERs) vary under different measurement bases in the experiment, we perform the following operations to improve the calculated secure key rate according to Eq.~\eqref{eq:1}. First, we define the transformations: $\ket{H \equiv +}$, $\ket{V \equiv -}$, $\ket{D \equiv 0}$ and $\ket{A \equiv 1}$, where $\ket{D}=(\ket{H}+\ket{V})/\sqrt{2}$ and $\ket{A}=(\ket{H}-\ket{V})/\sqrt{2}$ represent the diagonal and anti-diagonal polarization states, respectively. Then, after recording the experimental measurement results using the encoding scheme, we perform a classical post-processing operation equivalent to the quantum operation $X_AZ_A \otimes I_{B_i}$ on the acquired data. Here, $X$ and $Z$ denote the Pauli operators, $I$ represents the identity operator, and the subscripts indicate their respective target ($A$ for Alice and $B_i$ for $\text{Bob}_i$). After these operations, the quantum states of the qubits can be equivalently expressed as $\ket{\Phi^+} = (\ket{00} + \ket{11})/\sqrt{2}$ in the $Z$ basis and  $\ket{\Phi^+} = (\ket{++} + \ket{--})/\sqrt{2}$ in the $X$ basis.
This conversion improves the secure key rate and ensures the consistency between the experimental system and the theoretical protocol. To simplify the subsequent descriptions, we treat the measured quantum states as $\ket{\Phi^+}$ in the following discussions.

To investigate the influence of the $Z$-basis selection probability $p_z$ and the channel transmission on the secure key rate, we select probabilities of 0.9 and 0.5, and conduct the experiment under three different channel transmission values, thereby calculating the secure key rates for a total of six scenarios with Eq.~\eqref{eq:1}. The initial channel transmission values between each user and the central node Eve are all measured to be about 1.64 $\times$ $10^{-1}$. In experiments, the losses of all the VOAs are sequentially set to 0 dB, 3 dB and 5 dB to change the channel transmission. The accumulation time is approximately 1034 seconds for each scenario, during which a total of $10^{11}$ pulses are sent. The average down-converted mean photon number  is maintained at $\mu = 0.023$. The QBERs and secure key rates are calculated from the experimental data. The detailed results are summarized in Table~\ref{tab:1}. The upper limit of the phase error rate $\phi^Z$ can be estimated based on the total bit error rate in the $X$ basis, which can be inferred from the bit error rate between Alice and $\text{Bob}_i$ in the $X$ basis ($E_{AB_i}^X$)~\cite{bao2024efficient}. Therefore, the QBERs in $Z$ basis and $X$ basis are both provided. 

\begin{figure}[htbp]
    \centering
    \includegraphics[width=0.7\linewidth]{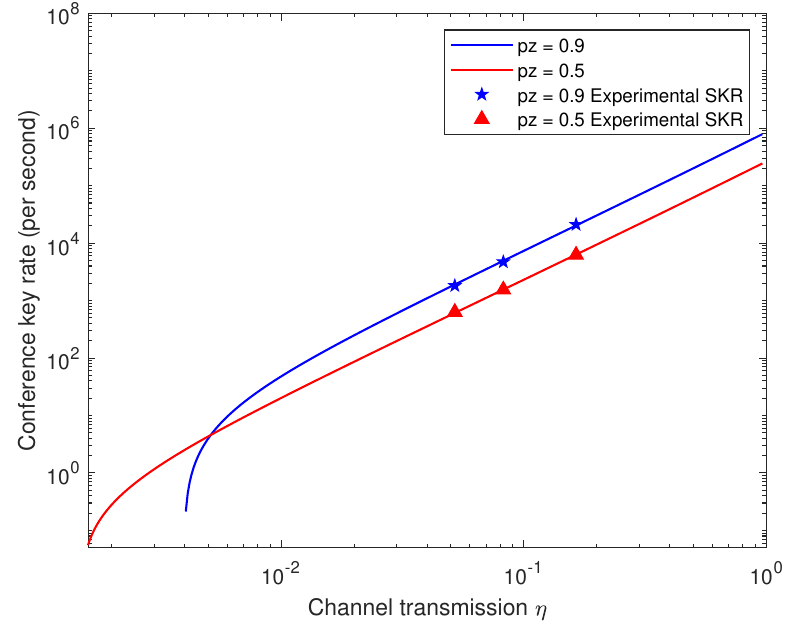}
    \caption{Conference secure key rate as a function of channel transmission $\eta$. The five-pointed blue stars and red triangles represent the experimental results for $p_z = 0.9$ and $p_z = 0.5$, respectively. For each $p_z$, the QCKA key rates are measured under three different channel transmission values, where the channel transmission $\eta$ denotes the transmission from the central node to each user. The blue and red lines represent the simulation results for $p_z = 0.9$ and $p_z = 0.5$, respectively.}
    \label{fig4}
\end{figure}

\begin{table*}[!t]
    \caption{ The parameters involved in the secure key rate calculation. $e_0$ is the background error rate. $e_d$ is the misalignment error rate. $\eta_d$ is the detection efficiency of single photon detectors. $p_d$ is the dark count rate per pulse. $f$ is the error correction efficiency. $\epsilon_{\mathrm{cor}}$ and $\epsilon_{\mathrm{sec}}$ represent the parameters of correctness and privacy, respectively.}
    \begin{tabular*}{\hsize}{@{}@{\extracolsep{\fill}}lllllll@{} }
    \toprule
    $e_0$ &  $e_d$ & $\eta_d$ & $p_d$ & $f$ & $\epsilon_{\mathrm{cor}}$ & $\epsilon_{\mathrm{sec}}$\\
    \midrule
        0.5 & 0.02 & 83\% & $10^{-7}$  & 1.16 & $1.2 \times 10^{-9}$ & $1.2 \times 10^{-9}$ \\
    \bottomrule
    \end{tabular*}
    \label{tab:2}
\end{table*}

Fig.~\ref{fig4} depicts the conference key rate versus channel transmission. The maximum secure key rate of $2.11 \times 10^{4}$ bit/s is attained at the channel transmission of 1.64 $\times$ $10^{-1}$ with a $Z$-basis selection probability of 0.9. At identical channel transmission, the key rate is markedly higher in most cases for $p_z = 0.9$ compared to $p_z = 0.5$. This is because the raw keys are generated from coincidence counts in the $Z$ basis, and intuitively, a higher $p_z$ corresponds to an increased key rate. However, as the channel transmission decreases, the key rate when $p_z = 0.9$ decreases more rapidly than when $p_z = 0.5$. A higher $p_z$ results in an increased number of coincidence counts in the $Z$ basis but a reduced number in the $X$ basis, thereby amplifying statistical uncertainties in the $X$-basis measurements, affecting the estimation of $\phi^Z$ and consequently decreasing the final secure key rate. With decreasing channel transmission, the experimental secure key rate remains at a high level, demonstrating the high efficiency and practicality of the protocol in quantum network applications.

\section{Discussion}
In summary, we have experimentally demonstrated the tripartite source-independent QCKA protocol. We establish a tripartite QCKA system, leveraging polarization-entangled photon pairs to facilitate secure multi-user communication. The experimental system is capable of generating high-quality entangled photon pairs with the fidelity of approximately $97\%$ and the visibility exceeding $96\%$. Notably, we achieve a high key rate of $2.11 \times 10^4$ bits/s at the channel transmission of 1.64 $\times$ $10^{-1}$ and a $Z$-basis selection probability of 0.9.

Furthermore, we conduct six sets of experiments to explore the effects of basis selection probability and channel transmission on the secure key rate. In most scenarios, the key rate for $p_z = 0.9$ is markedly higher than for $p_z = 0.5$. Nevertheless, as channel transmission decreases, the key rate when $p_z = 0.9$ declines more rapidly, which results from the increasing statistical fluctuation of coincidence counts in the $X$ basis. Additionally, the intensity of the source is directly correlated with the data size. While higher intensity increases coincidence counts, it also leads to higher QBERs. Therefore, it's crucial to choose an appropriate basis selection probability to optimize the key rate according to the channel transmission and the intensity of the source.  

Compared to previous QCKA experiments, our approach is more efficient and scalable based on entangled photon pairs, thereby offering a feasible route for implementing multipartite QCKA in large-scale quantum networks. By incorporating technologies such as wavelength division multiplexing, the implementation can be integrated into quantum networks,  including fully connected QKD networks, to accommodate larger user groups.  Overall, this work  marks a pivotal step toward the realization of an efficient, multi-user quantum network for next-generation secure communication systems.

\section{Materials and Methods}
\subsection{Supplementary Data}
\label{sec:4.1}
In this work, we have experimentally implemented the tripartite source-independent QCKA protocol, utilizing polarization-entangled photon pairs to enable scalable and secure multi-user communication.  A high key rate of  $2.11 \times 10^4$ bits/s is achieved with a compact experimental setup under the channel transmission of 1.64 $\times$ $10^{-1}$. The parameters used in the calculation of the secure key rates are provided in Table~\ref{tab:2}. $e_0$ is the background error rate. $e_d$ is the misalignment error rate. $\eta_d$ is the detection efficiency of single photon detectors. $p_d$ is the dark count rate per pulse. $f$ is the error correction efficiency. $\epsilon_{\mathrm{cor}}$ and $\epsilon_{\mathrm{sec}}$ are the parameters of correctness and privacy. In the simulation of theoretical predictions, the parameters remain the same.

During the experiment, a total of $10^{11}$ pulses are sent. The detailed coincidence counts in the $Z$ basis and $X$ basis under varying channel transmission and key selection probabilities are presented in Table~\ref{tab:3} and Table~\ref{tab:4}. Table~\ref{tab:3} presents the coincidence counts between Alice and $\text{Bob}_1$ and Table~\ref{tab:4} provides the coincidence counts between Alice and $\text{Bob}_2$. The state of the detected photon pairs is expressed as $\ket{\Phi^+}=(\ket{00}+\ket{11})/\sqrt{2}$ in the $Z$ basis and $\ket{\Phi^+}=(\ket{++}+\ket{--})/\sqrt{2}$ in the $X$ basis. Notably, the theoretical prediction reveals an optimal basis selection probability, which is dependent on the down-converted mean photon number $\mu$ and the channel transmission. However, it is challenging for us to achieve the optimal ratio due to constraints imposed by the splitting ratios of BSs. Thus, there remains potential for further enhancement in the key rates. Through the coincidence counts, we can obtain the number of entangled photon pairs detected in the $Z$ basis $n_Z$ and calculate the QBERs in the $Z$ basis and $X$ basis, which are essential for estimating the upper bound of phase error rate in the $Z$ basis $\phi^Z$ and the length of the secure key $L_{\mathrm{QCKA}}$. The down-converted mean photon numbers $\mu$ of the two entangled photon pairs are different due to the difficulty in achieving perfect splitting of laser pulses.

\subsection{Computation of the Four Expectations to Calculate S parameter}
To calculate the S parameter of the CHSH inequality, we need to calculate the normalized expectation value $E(\theta_A, \theta_B)$ based on the results of polarization measurements. The definition of the expectation is:
\begin{equation}
    E(\theta_A, \theta_B)= \frac{N_{++}-N_{+-}-N_{-+}+N_{--}}{N_{++}+N_{+-}+N_{-+}+N_{--}}
\end{equation}
where $N_{ij}$ represents the coincidence counts ($+$ denotes parallel alignment, $-$ denotes antiparallel alignment of the angle of the two users). The experimental results are presented in Fig.~\ref{fig3}. To obtain $E(\theta_A, \theta_B)$, Alice and $\text{Bob}_i$ adjust the angle of their polarizer to measure the coincidence counts under the four scenarios corresponding to parallel and antiparallel alignments with a determined initial angle combination $(\theta_A, \theta_B)$. The average coincidence counts measured in the experiment are shown in Table~\ref{tab:5} and Table~\ref{tab:6}.

\begin{table*}[!t]
    \caption{The detailed coincidence counts under different channel transmission and key selection probabilities between Alice and $\text{Bob}_1$. $\eta$ denotes the channel transmission excluding the detector efficiency between central node Eve and each user. The $Z$ ($X$) indicates that Alice and $\text{Bob}_1$ select the same $Z$ basis ($X$ basis). The subscript (00, 01, 10, 11) correspond to the measurement results of the detected photon pair. The down-converted mean photon number $\mu$ is approximately 0.024 per pulse.}
    \begin{tabular*}{\hsize}{@{}@{\extracolsep{\fill}}llll|lll@{} }
    \hline
    $p_z$ &  &0.9 &  & & 0.5 &  \\
    \hline
    $\eta$  & 1.64 $\times$ $10^{-1}$ & 8.24 $\times$ $10^{-2}$ & 5.20 $\times$ $10^{-2}$ & 1.64 $\times$ $10^{-1}$ & 8.24 $\times$ $10^{-2}$ & 5.20 $\times$ $10^{-2}$ \\
    \hline
    $Z_{00}$ & 17559115  & 4159589  & 1610121  & 5243818  & 1376112  & 563083   \\ 
    $Z_{01}$ & 520307  & 159452  & 64587  & 91583  & 23690  & 10013   \\ 
    $Z_{10}$ & 388668  & 99534  & 41610  & 99125  & 31773  & 12752   \\ 
    $Z_{11}$ & 17193668  & 4144120  & 1586726  & 5247370  & 1350268  & 558092   \\ 
    $X_{00}$ & 119561  & 29056  & 12096  & 5670628  & 1418979  & 588371   \\ 
    $X_{01}$ & 1774  & 426  & 198  & 129711  & 33365  & 13971   \\ 
    $X_{10}$ & 1469  & 397  & 190  & 83117  & 22982  & 10234   \\ 
    $X_{11}$ & 99272  & 25894  & 11553  & 4797483  & 1269066  & 535394   \\
    \hline
    \end{tabular*}
    \label{tab:3}
\end{table*}

\begin{table*}[!t]
    \caption{The detailed coincidence counts under different channel transmission and key selection probabilities between Alice and $\text{Bob}_2$. $\eta$ denotes the channel transmission excluding the detector efficiency between central node Eve and each user. The $Z$ ($X$) means that Alice and $\text{Bob}_2$ choose the same $Z$ basis ($X$ basis). The superscript (00, 01, 10, 11) correspond to the measurement results of the photon pair. The down-converted mean photon number $\mu$ is approximately 0.021 per pulse.}
    \begin{tabular*}{\hsize}
    {@{}@{\extracolsep{\fill}}llll|lll@{} }
    \hline
    $p_z$ &  &0.9 &  & & 0.5 &  \\
    \hline
    $\eta$  & 1.64 $\times$ $10^{-1}$ & 8.24 $\times$ $10^{-2}$ & 5.20 $\times$ $10^{-2}$ & 1.64 $\times$ $10^{-1}$ & 8.24 $\times$ $10^{-2}$ & 5.20 $\times$ $10^{-2}$ \\
    \hline
        $Z_{00}$ & 28787736  & 6667492  & 2749255  & 7156545  & 1743702  & 731003   \\ 
        $Z_{01}$ & 885955  & 200731  & 48101  & 165071  & 40469  & 14390   \\ 
        $Z_{10}$ & 632049  & 115341  & 79320  & 108417  & 33147  & 17489   \\ 
        $Z_{11}$ & 26663254  & 6443307  & 2569273  & 6872625  & 1729300  & 755435   \\ 
        $X_{00}$ & 182153  & 44402  & 17940  & 7742444  & 1930379  & 688314   \\ 
        $X_{01}$ & 2348  & 621  & 259  & 141563  & 35871  & 15346   \\ 
        $X_{10}$ & 1680  & 561  & 197  & 152882  & 48212  & 13195   \\ 
        $X_{11}$ & 182327  & 53928  & 18482  & 7281136  & 1842162  & 652107   \\ 
        \hline
    \end{tabular*}
    \label{tab:4}
\end{table*}

\begin{table*}[!t]
    \renewcommand{\arraystretch}{1.2}
    \caption{The average coincidence counts $N$ measured under different combination of the polarization angles of Alice and $\text{Bob}_1$.}
    \centering
    \newcolumntype{c}{>{\centering\arraybackslash}X}
    \begin{tabularx}{\textwidth}{c|c|c|c|c}
    \hline
    $\left(\theta_A, \theta_B\right)$ & 
    $\left(0^{\circ}, -22.5^{\circ}\right)$ & 
    $\left(0^{\circ}, 22.5^{\circ}\right)$ & 
    $\left(-45^{\circ}, -22.5^{\circ}\right)$ & 
    $\left(-45^{\circ}, 22.5^{\circ}\right)$ \\
    \hline
    $N_{++}$ & 15310 & 11512 & 12883 & 66602  \\ 
    \hline
    $N_{+-}$ & 68214 & 73735 & 73433 & 15465  \\ 
    \hline
    $N_{-+}$ & 68063 & 71460 & 72430 & 15258  \\ 
    \hline
    $N_{--}$ & 15518 & 10817 & 10585 & 67398  \\ 
    \hline
    \end{tabularx}
    \label{tab:5}
\end{table*}

\begin{table*}[!t]
    \renewcommand{\arraystretch}{1.2}
    \caption{The average coincidence counts $N$ measured under different combination of the polarization angles of Alice and $\text{Bob}_2$.}
    \newcolumntype{c}{>{\centering\arraybackslash}X}
    \begin{tabularx}{\textwidth}{c|c|c|c|c}
    \hline
    $\left(\theta_A, \theta_B\right)$ & 
    $\left(0^{\circ}, -22.5^{\circ}\right)$ & 
    $\left(0^{\circ}, 22.5^{\circ}\right)$ & 
    $\left(-45^{\circ}, -22.5^{\circ}\right)$ & 
    $\left(-45^{\circ}, 22.5^{\circ}\right)$ \\
    \hline
    $N_{++}$ & 17035 & 16706 & 23747 & 99819  \\ 
    \hline
    $N_{+-}$ & 98111 & 101814 & 96475 & 19172  \\ 
    \hline
    $N_{-+}$ & 106584 & 103258 & 101611 & 20995  \\ 
    \hline
    $N_{--}$ & 17961 & 19506 & 18746 & 100198  \\ 
    \hline
    \end{tabularx}
    \label{tab:6}
\end{table*}

\subsection{Density matrix reconstruction via maximum likelihood estimation}

To ensure the reconstructed density matrix is physical, i.e., Hermitian, positive semi-definite, and with unit trace, a maximum likelihood estimation (MLE) method is employed~\cite{james2001measurement}. This approach circumvents the issues of linear tomography, which can yield non-physical matrices due to experimental noise.

The MLE procedure begins by constructing an explicitly physical density matrix, $\hat{\rho}_{p}$. First, we define an arbitrary $4 \times 4$ lower-triangular matrix $\hat{T}(t)$, which is a function of 16 independent real parameters $\{t_1, t_2, \dots, t_{16}\}$. A matrix $\hat{\mathcal{G}} = \hat{T}^{\dagger}\hat{T}$ is necessarily Hermitian and positive semi-definite. By normalizing this matrix, a valid physical density matrix is constructed as:
\begin{equation}
    \hat{\rho}_{p}(t) = \frac{\hat{T}^{\dagger}(t)\hat{T}(t)}{\Tr\{\hat{T}^{\dagger}(t)\hat{T}(t)\}}
\end{equation}

With this representation, the goal of the MLE method is to find the optimal set of parameters $t$ that best describes the experimental data. This is achieved by defining a likelihood function, $\mathcal{L}(t)$, which quantifies the differences between the measured coincidence counts $n_{\nu}$ and the expected counts $\overline{n}_{\nu}(t)$ predicted by the physical density matrix $\hat{\rho}_{p}(t)$. Here, $\nu = 1, 2, \dots, 16$ labels the distinct measurement settings. 
The expected count $\overline{n}_{\nu}(t)$ is defined as:
\begin{equation}
    \overline{n}_{\nu}(t) = \mathcal{N}\bra{\psi_{\nu}}\hat{\rho}_{p}(t)\ket{\psi_{\nu}}
\end{equation}
where $\mathcal{N}$ is the normalization constant representing the total number of events for a complete basis. $\mathcal{N}$ is estimated from the experimental data as $\mathcal{N} = n_1 + n_2 + n_3 + n_4$. $\ket{\psi_{\nu}}$ denotes the quantum state vector representing the $\nu$-th tomographic projection measurement. For example, as defined in Table \ref{tab:7}, $\ket{\psi_1} = \ket{HH}$, $\ket{\psi_2} = \ket{HV}$, etc. 

Assuming Gaussian probability distribution for the count noise, the optimization problem reduces to finding the minimum of the function $\mathcal{L}(t)$:
\begin{equation}
    \mathcal{L}(t) = \sum_{\nu=1}^{16} \frac{[\overline{n}_{\nu}(t) - n_{\nu}]^2}{2\overline{n}_{\nu}(t)} = \sum_{\nu=1}^{16} \frac{[\mathcal{N}\bra{\psi_{\nu}}\hat{\rho}_{p}(t)\ket{\psi_{\nu}} - n_{\nu}]^2}{2\mathcal{N}\bra{\psi_{\nu}}\hat{\rho}_{p}(t)\ket{\psi_{\nu}}} 
    \label{eq:likelihood}
\end{equation}
A numerical optimization algorithm is used to find the specific set of parameters, $\{t_{1}^{(\mathrm{opt})}, \dots, t_{16}^{(\mathrm{opt})}\}$, that minimizes this function $\mathcal{L}$. The resulting matrix, $\hat{\rho}_{p}(t^{(\mathrm{opt})})$, is then taken as the most likely physical density matrix consistent with the observed data. The coincidence counts $n_\nu$ measured in the experiment and the corresponding wave plate angles $\{q_{1,\nu},h_{1,\nu},q_{2,\nu},h_{2,\nu}\}$ are provided in Table \ref{tab:7} and Table \ref{tab:8}. To reduce statistical fluctuations, the coincidence counts $n_\nu$ for each measurement setting are accumulated for 3 seconds. The notation used for the polarization states is $\ket{D} \equiv (\ket{H}+\ket{V})/\sqrt{2}$ and $\ket{R} \equiv (\ket{H}-i\ket{V})/\sqrt{2}$.

\begin{table*}[ht]
\centering
\caption{Experimental coincidence counts ($n_\nu$) measured under the 16 tomographic projection settings for Alice and Bob$_1$.}
\label{tab:7}
\begin{tabular}{cccrrrrrr}
\toprule
$\nu$ & Mode 1 & Mode 2 & \multicolumn{1}{c}{$q_1$} & \multicolumn{1}{c}{$h_1$} & \multicolumn{1}{c}{$q_2$} & \multicolumn{1}{c}{$h_2$} & \multicolumn{1}{c}{$n_\nu$} \\
\midrule
1 & $\ket{H}$ & $\ket{H}$ & $0^{\circ}$ & $0^{\circ}$ & $0^{\circ}$ & $0^{\circ}$ & 1305 \\
2 & $\ket{H}$ & $\ket{V}$ & $0^{\circ}$ & $0^{\circ}$ & $0^{\circ}$ & $45^{\circ}$ & 82190 \\
3 & $\ket{V}$ & $\ket{V}$ & $0^{\circ}$ & $45^{\circ}$ & $0^{\circ}$ & $45^{\circ}$ & 1257 \\
4 & $\ket{V}$ & $\ket{H}$ & $0^{\circ}$ & $45^{\circ}$ & $0^{\circ}$ & $0^{\circ}$ & 81980 \\
5 & $\ket{R}$ & $\ket{V}$ & $0^{\circ}$ & $22.5^{\circ}$ & $0^{\circ}$ & $45^{\circ}$ & 36899 \\
6 & $\ket{R}$ & $\ket{H}$ & $0^{\circ}$ & $22.5^{\circ}$ & $0^{\circ}$ & $0^{\circ}$ & 43677 \\
7 & $\ket{D}$ & $\ket{H}$ & $45^{\circ}$ & $22.5^{\circ}$ & $0^{\circ}$ & $0^{\circ}$ & 48880 \\
8 & $\ket{D}$ & $\ket{V}$ & $45^{\circ}$ & $22.5^{\circ}$ & $0^{\circ}$ & $45^{\circ}$ & 37391 \\
9 & $\ket{D}$ & $\ket{R}$ & $45^{\circ}$ & $22.5^{\circ}$ & $0^{\circ}$ & $22.5^{\circ}$ & 44333 \\
10 & $\ket{D}$ & $\ket{D}$ & $45^{\circ}$ & $22.5^{\circ}$ & $45^{\circ}$ & $22.5^{\circ}$ & 1585 \\
11 & $\ket{R}$ & $\ket{D}$ & $0^{\circ}$ & $22.5^{\circ}$ & $45^{\circ}$ & $22.5^{\circ}$ & 47749 \\
12 & $\ket{R}$ & $\ket{R}$ & $0^{\circ}$ & $22.5^{\circ}$ & $0^{\circ}$ & $22.5^{\circ}$ & 1863 \\
13 & $\ket{H}$ & $\ket{D}$ & $0^{\circ}$ & $0^{\circ}$ & $45^{\circ}$ & $22.5^{\circ}$ & 48302 \\
14 & $\ket{H}$ & $\ket{R}$ & $0^{\circ}$ & $0^{\circ}$ & $0^{\circ}$ & $22.5^{\circ}$ & 38370 \\
15 & $\ket{V}$ & $\ket{R}$ & $0^{\circ}$ & $45^{\circ}$ & $0^{\circ}$ & $22.5^{\circ}$ & 46510 \\
16 & $\ket{V}$ & $\ket{D}$ & $0^{\circ}$ & $45^{\circ}$ & $45^{\circ}$ & $22.5^{\circ}$ & 40014 \\
\bottomrule
\end{tabular}
\end{table*}

\begin{table*}[!t]
\centering
\caption{Experimental coincidence counts ($n_\nu$) measured under the 16 tomographic projection settings for Alice and Bob$_2$.}
\label{tab:8}
\begin{tabular}{cccrrrrrr}
\toprule
$\nu$ & Mode 1 & Mode 2 & \multicolumn{1}{c}{$q_1$} & \multicolumn{1}{c}{$h_1$} & \multicolumn{1}{c}{$q_2$} & \multicolumn{1}{c}{$h_2$} & \multicolumn{1}{c}{$n_\nu$} \\
\midrule
1 & $\ket{H}$ & $\ket{H}$ & $0^{\circ}$ & $0^{\circ}$ & $0^{\circ}$ & $0^{\circ}$ & 1383 \\
2 & $\ket{H}$ & $\ket{V}$ & $0^{\circ}$ & $0^{\circ}$ & $0^{\circ}$ & $45^{\circ}$ & 119018 \\
3 & $\ket{V}$ & $\ket{V}$ & $0^{\circ}$ & $45^{\circ}$ & $0^{\circ}$ & $45^{\circ}$ & 1704 \\
4 & $\ket{V}$ & $\ket{H}$ & $0^{\circ}$ & $45^{\circ}$ & $0^{\circ}$ & $0^{\circ}$ & 121674 \\
5 & $\ket{R}$ & $\ket{V}$ & $0^{\circ}$ & $22.5^{\circ}$ & $0^{\circ}$ & $45^{\circ}$ & 60747 \\
6 & $\ket{R}$ & $\ket{H}$ & $0^{\circ}$ & $22.5^{\circ}$ & $0^{\circ}$ & $0^{\circ}$ & 62533 \\
7 & $\ket{D}$ & $\ket{H}$ & $45^{\circ}$ & $22.5^{\circ}$ & $0^{\circ}$ & $0^{\circ}$ & 73073 \\
8 & $\ket{D}$ & $\ket{V}$ & $45^{\circ}$ & $22.5^{\circ}$ & $0^{\circ}$ & $45^{\circ}$ & 50917 \\
9 & $\ket{D}$ & $\ket{R}$ & $45^{\circ}$ & $22.5^{\circ}$ & $0^{\circ}$ & $22.5^{\circ}$ & 77260 \\
10 & $\ket{D}$ & $\ket{D}$ & $45^{\circ}$ & $22.5^{\circ}$ & $45^{\circ}$ & $22.5^{\circ}$ & 3037 \\
11 & $\ket{R}$ & $\ket{D}$ & $0^{\circ}$ & $22.5^{\circ}$ & $45^{\circ}$ & $22.5^{\circ}$ & 57551 \\
12 & $\ket{R}$ & $\ket{R}$ & $0^{\circ}$ & $22.5^{\circ}$ & $0^{\circ}$ & $22.5^{\circ}$ & 2585 \\
13 & $\ket{H}$ & $\ket{D}$ & $0^{\circ}$ & $0^{\circ}$ & $45^{\circ}$ & $22.5^{\circ}$ & 65648 \\
14 & $\ket{H}$ & $\ket{R}$ & $0^{\circ}$ & $0^{\circ}$ & $0^{\circ}$ & $22.5^{\circ}$ & 58292 \\
15 & $\ket{V}$ & $\ket{R}$ & $0^{\circ}$ & $45^{\circ}$ & $0^{\circ}$ & $22.5^{\circ}$ & 60181 \\
16 & $\ket{V}$ & $\ket{D}$ & $0^{\circ}$ & $45^{\circ}$ & $45^{\circ}$ & $22.5^{\circ}$ & 61324 \\
\bottomrule
\end{tabular}
\end{table*}

\section*{Acknowledgments}

\subsection*{Author Contributions} 
Z.-B.C. guided the work. H.-L.Y. conceived and supervised the research. W.-J.H., Y.-R.X., Y.B., and H.-L.Y. designed the experiments. W.-J.H., Y.-R.X., and H.-L.Y. performed the experiments and analyzed the experimental data. W.-J.H., Y.-R.X., and H.-L.Y. cowrote the manuscript, with input from the other authors. All authors have discussed the results and proofread the manuscript.

\subsection*{Funding}
This work was supported by the National Natural Science Foundation of China (Nos. 12522419, U25D8016, and 
12274223), the Fundamental Research Funds for the Central Universities and the Research Funds of Renmin University of China (No. 24XNKJ14).

\subsection*{Conflicts of Interest}
The authors declare that they have no competing interests.

\subsection*{Data Availability}
All data that support the findings of this study are available from the corresponding authors upon reasonable request. 


\end{document}